\documentclass[twocolumn]{revtex4-1}
\usepackage{graphicx}
\usepackage{srcltx,amssymb,color,float}
\usepackage{hyperref}
\usepackage{cleveref}
\usepackage{multirow}

\def\ep{\varepsilon}
\def\pa{\partial}

\def\fmc{\rm fm^{-3}} 
\def\fm{\rm fm}
      
\newcommand{\beq}{\begin{equation}}
\newcommand{\eeq}{\end{equation}}
\newcommand{\beqa}{\begin{eqnarray}}
\newcommand{\eeqa}{\end{eqnarray}}
\newcommand {\apgt} {\ {\raise-.5ex\hbox{$\buildrel>\over\sim$}}\ }

\newcommand{\der}[2]{\frac{\partial #1}{\partial #2}}
\newcommand{\derc}[3]{\left( \frac{\partial #1}{\partial #2} \right)
 \raisebox{-1em}{\ensuremath{#3}}}

\newcommand{\vecb}[1]{\mbox{\boldmath $#1$}}

\graphicspath{{figs/}}

\begin{document}

\title{Exact solution of equations for proton localization in neutron star 
matter}
\author{Sebastian~Kubis}
\email{skubis@pk.edu.pl}
\author{W\l{}odzimierz~W\'ojcik}
\address{Institute of Physics, Cracow University of Technology, Podchor\c{a}\.zych 1, 30-084 Krak\'ow, Poland}
\begin{abstract}

The rigorous treatment of proton localization phenomenon in asymmetric nuclear
matter is presented.  The solution of  proton wave function  and neutron
background distribution is found by the use of the extended Thomas-Fermi
approach. The minimum of energy is obtained in the Wigner-Seitz approximation of
spherically symmetric cell. The analysis of three different nuclear models 
suggests  that the proton localization is likely to take place in the interior
of neutron star.
\vskip0.5cm
PACS number(s): 21.65.Cd, 21.65.Mn, 26.60.-c
\end{abstract}

\maketitle
\section{Introduction}

The interior of a neutron star contains the densest forms of matter in the
Universe. The central density is as high as 5 to 10 times the nuclear
equilibrium density $n_0=0.16 ~\fmc$.  Most of the mass of the star is placed in
its liquid core covered by a thin crust ($< 1 \rm~km$ for typical neutron star)
whose bottom edge is located at around 0.5$n_0$.  Above this density the
matter is well described by the Fermi liquid - a mixture of nucleons and
leptons. In comparison to the matter  present inside the stable nuclei, the
matter in neutron star is highly asymmetric as a consequence of the 
$\beta$-equilibrium taking place between nucleons and leptons. It is convenient to
express the asymmetry by the proton fraction $x=n_p/n$, where $n_p, n$ are the
proton and baryon number density. The proton fraction is between 0.4 and 0.5  in
nuclei, whereas in  neutron star matter at $n_0$ it is equal to  4\% what is
exactly determined by the saturation point  properties. The proton abundance at
higher density is not well known and different nuclear interactions models lead
to a very large discrepancies in the $x(n)$ behaviour. There are models which
predict that $x$   does not
exceed 10\% in a full range of densities. When the proton fraction is not high,
protons can be treated as the  small admixture to the neutron background,  where
direct proton-proton interaction is negligible and hence protons can be regarded
as  impurities in the  neutron matter. Therefore, a description of this system
by single proton in neutron  background is justified. The  attractive nature of 
proton-neutron  interaction may result in an instability of  homogeneously
distributed protons
\cite{kutschera89,kutschera90,kutschera93}.
In the paper \cite{kutschera93} the polaron behaviour of a proton impurity in
dense neutron matter was discussed. A single proton in neutron matter can lower its
energy  by inducing the density inhomogeneity around it. 
  Instead of forming the Fermi sea,
protons occupy ground state with zero momentum above some critical density. 
It occurs when the  localized proton with properly distorted  neutron background
has smaller energy than the system with the proton described by the 
plane waves. Such a state of
matter has intriguing magnetic properties that has been shown in
\cite{kutschera89,kutschera94,kutschera97}. It exhibits, e.g. a crystallization
of proton impurities in the neutron star interior \cite{kutschera95,potekhin10} 
and affects the cooling process of neutron stars \cite{baiko99,baiko00}. In the
papers  \cite{kutschera90,kutschera93} the proton localization has been
analyzed  in a simplified manner. A variational approach to a cell containing 
one proton was proposed. However, the minimization of the energy was achieved 
with respect to Gaussian-type trial  function with only one parameter 
for both proton wave function and neutron background. Moreover, the cell was
treated as a system with infinite  volume $V\rightarrow \infty$ which means that
the method is applicable to a very small ($x$ smaller than 1\%) proton fraction.

The aim of this work is to solve exactly the Lagrange-Euler equations
corresponding to the variational approach proposed in the original works. We
also abandon the  assumption of infinitely large  cell.  This
means we may take  into account  higher proton fractions and thus, extent 
the class of nuclear models in the analysis. 

\begin{figure}
\includegraphics[clip,width=\columnwidth]{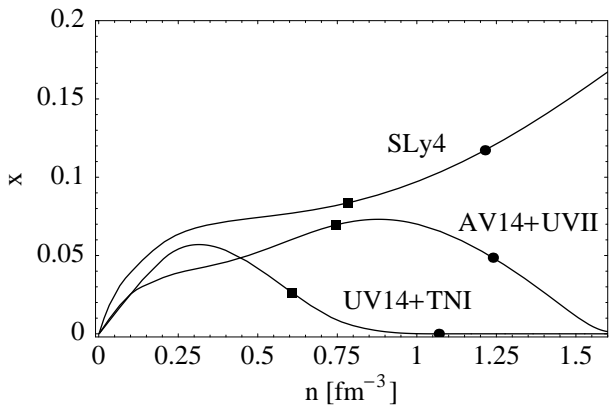}

\caption{The proton fraction for different models used in the calculation:
AV14+UVII, SLy4 and UV14+TNI. The squares indicate the proton localization
threshold and the full dots indicate the central density of
a star with maximum mass.}
\label{x.eps}
\end{figure}

  The equation of state of supra-nuclear density in a neutron star core  cannot
be  calculated unambiguously \cite{lattimer07,haensel06}. Instead, there are
many theoretical  models of exotic matter.  We choose two representatives of
them satisfying the criterion of maximum mass greater then $2M_\odot$
\cite{maxmass1,maxmass2} and leading, at the  same time, to not very high
proton  fraction at higher densities:   AV14+UVII \cite{wff88}, SLy4
\cite{chabanat98}.  The last one is very common in the description of the
neutron stars as it well reproduces the  properties of nuclei, the 
nucleon-nucleon scattering data, and it correctly recovers saturation point 
properties. Moreover, in our analysis we also included UV14+TNI model taken from
\cite{lagaris81}. Although the maximum neutron star mass for this model is
smaller than $2M_\odot$, it was interesting to test it for 
very small proton fractions at high density because it could be relevant 
for the
properties of  proton localization.  In the Fig.\ref{x.eps} the proton fraction
in  $\beta$-equilibrated matter is shown for the three selected nuclear models.

The present paper is organized as follows.  A short review of the  the variational
method for finite-size Wigner-Seitz cell is presented in Sec.II.  In Sec.III
 the numerical method for solving the equations is explained. The results are
shown and discussed for various nuclear models in Sec.IV.

\vskip2.0cm

\section{A proton in neutron background}

In order to calculate the energy of nuclear matter with localized protons 
we treat the proton as a quantum particle described by its wave function
$\Psi(\vecb{r})$ whereas the neutrons are represented  by a density
distribution function $n_n(\vecb{r})$. Like in the work 
\cite{kutschera93} we assume
that one proton occupies a spherical Wigner-Seitz (W-S) cell filled with large number
of neutrons. Neutrons are treated in the local density approximation
 according to differential Thomas-Fermi scheme \cite{bbp71}. 
The energy of the cell 
 is expressed by the integral over the whole cell volume $V_{WS}=1/\bar{n}_p$
\beqa
E[\psi,n_n] &= {\displaystyle \int_{V_{WS}}} &\left[ \Psi^*(- \frac{\nabla^2}{2 m_p}+ 
\mu_p)\Psi \right. \nonumber \\ 
 & &\Big. + \; \ep + B_N (\nabla n_n)^2\Big] d^3 r .
\label{Ews}
\eeqa

The nuclear matter energy density $\ep$  is the thermodynamical function which
depends directly on nucleon densities $\ep(n_n,n_p)$. Its functional form  is
completely determined by the adopted nuclear model. The chemical potentials are
defined as  usual
\beq
 \mu_p = \derc{\ep}{n_p}{n_n}~~,~~  \mu_n =
\derc{\ep}{n_n}{n_p} .
\label{mi}
\eeq
 In the energy functional  Eq.~(\ref{Ews}) the energy density $\ep$
and the proton chemical potential $\mu_p$ get the  space dependence
 by the local 
neutron density: $\ep(r)=\ep(n_n(r),0)$ and in the same
way $\mu_p(r)=\mu_p(n_n(r),0)$. The constant coefficient
$B_N$ describes the gradient contribution and it is fitted to 
the surface properties of nuclei, here we
adopt the value $B_N=31.6~ \rm MeV\; fm^5  $  \cite{kutschera90} .

The W-S cell radius is given by the proton density for homogeneous system
$R_{WS}=(3/4\pi \bar{n}_p)^{1/3}$, where $\bar{n}_p = x n$ and $n$ 
is the mean baryon number.

The cell energy should be minimized under constraints of fixed proton and
neutron number
\beqa
&{\displaystyle \int_{V_{WS}}} &\Psi^* \Psi d^3r = 1 \label{constraintpsi} ,\\
&{\displaystyle \int_{V_{WS}}} &n_n d^3r = V_{WS}\bar{n}_n \label{constraintnn},
\eeqa
where $\bar{n}_n = (1-x)n$ is the mean neutron number in the case of homogeneous
system. The constraints expressed by
Eqs.~(\ref{constraintpsi},\ref{constraintnn}) require the following 
Lagrange multipliers $\lambda_p, \lambda_n$:
\beq
\tilde{E} =  E - \lambda_p\int(\Psi^*\Psi - 1/V_{WS})d^3r -
 \lambda_n\int(n_n -\bar{n}_n) d^3r .
\eeq
For the isolated, spherically symmetric  W-S cell we 
impose  the following boundary conditions:
\beqa
\frac{\pa\Psi}{\pa r}(0) = 0 &~~,~~& \Psi(R_{WS}) = 0~,  \nonumber \\
\frac{\pa n_n}{\pa r}(0) = 0 &~~,~~& \frac{\pa  n_n}{\pa r}(R_{WS}) = 0 ~.
\eeqa
From the Lagrange-Euler equations for the minimum of $\tilde{E}$
one may remark that the Lagrange multipliers $\lambda_p$ and $\lambda_n$
correspond to the physical quantities such as
the eigenvalue $E_p$ of the proton wave
function and the neutron chemical potential $\mu_n$ at the cell boundary:
\beq
\lambda_p=E_p ~~,~~\lambda_n=\mu_n|_{R_{WS}},
\label{lambdas}
\eeq
and then, finally, one may write the Lagrange-Euler equations 
in the form 
\beq
\frac{- \nabla^2}{2 m_p}\Psi + \mu_p \Psi = E_p \Psi  ,
\label{eqpsi}
\eeq
\beq
\der{\mu_p}{n_n}\Psi^*\Psi + \Delta\mu_n   - 2B_N \nabla^2 n_n =0 ,
\label{eqnn}
\eeq
where $\Delta\mu_n = \mu_n(n_n(r),0) - \mu_n|_{R_{WS}}$ is the difference 
between local chemical potential 
and its boundary value.
The first equation Eq.~(\ref{eqpsi}) represents  the Schr\"odinger equation for 
the spherically symmetric proton wave function $\Psi(r)$
with the eigenvalue $E_p$. The coupling to neutron density $n_n(r)$ comes from
the chemical potential $\mu_p(n_n,0)$. 
The second equation Eq.~(\ref{eqnn}) is the nonlinear elliptic equation for the neutron density
distribution $n_n(r)$ coupled to the proton density $\Psi^*\Psi$.

  The proton localization occurs if at a given mean density
 $n$ there exists a proton wave function with a negative
eigenvalue $E_p<0$ 
  and  when the energy of homogeneous system of nucleons
is greater than the energy $E[\Psi,n_n]$ of system with distorted densities, 
that means
\beq
\Delta E = E[\Psi,n_n] - \ep(n(1\!-\!x),nx) V_{WS} <0  .
\label{DE}
\eeq
In this way, by solving the Eqs.~(\ref{eqpsi},\ref{eqnn}), we obtain a family of
solutions parametrized with the mean density of matter $n$. 

\vskip5.0cm

\section{The method}

The  mean baryon density $n$ does not enter directly  to the 
Eqs.~(\ref{eqpsi},\ref{eqnn}). The average density $n$ is determined indirectly by
the second constraint, Eq.(\ref{constraintnn}). Therefore, in numerical solving
it is simpler to set  the  value at the boundary
\beq
n_n^\infty \equiv
n_n|_{R_{WS}} ~,
\label{nninf}
\eeq
find the proton function and neutron background and then  finally  
derive the mean density from the relation

\beq
n = \frac{1}{V_{WS}}\left( 1 + \int_{V_{WS}} n_n(r) d^3 r \right)   ~.
\eeq

The set of Eqs.(\ref{eqpsi},\ref{eqnn}) was solved by the relaxation
method explained in the following.
\begin{figure}[t]
\includegraphics[clip,width=\columnwidth]{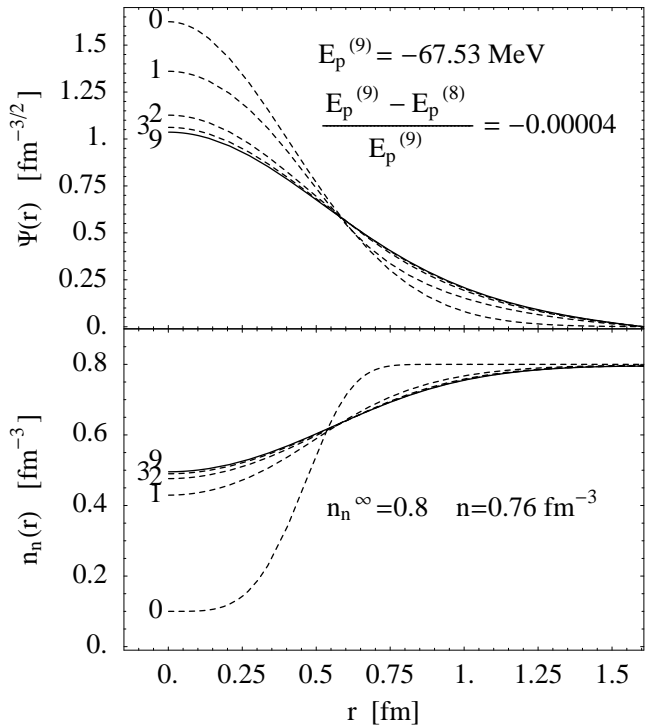}
\caption{The subsequent steps in the relaxation method for proton wave function
 $\psi(r)$ (upper panel) and neutron
density $n_n(r)$ (lower panel) in the
nuclear model AV14+UVII. The solid
curve represents the final results after  9 steps when the 
accuracy equal to $10^{-4}$ was achieved  for ground energy
$E_p$.}
\label{relax-psi-nn}
\end{figure}
As an initial approximation, the Gaussian-type function was taken 
for the proton wave function and for the neutron density.
In the $i$-th step, the iteration had two stages:
in the first  we found the ground state solution $\Psi^{(i+1)}~, E_p^{(i+1)}$ of 
Schr\"odinger equation including $n_n^{(i)}$. In the second step 
we solve the Eq.~(\ref{eqnn}) for $n_n^{(i+1)}$ including $\Psi^{(i+1)}$.
The iteration was continued  up to the point where  the 
eigenvalue does not change more than a given accuracy.
The procedure appeared to converge quickly, usually the accuracy equal
 to $10^{-4}$
was achieved in not more than 15 steps. 
The Fig.\ref{relax-psi-nn} represents the convergence of the iteration for the 
chosen density $n_n^\infty = 0.8$ in the  AV14+UVII model.

\section{Results}

The proton localization scheme  described in the previous sections was then
analyzed for  the three nuclear models: AV14+UVII, SLy4 and UV14+TNI. For
all of them the proton localization turned out to occur. Comparing the
localization threshold obtained here with the results of previous works based
 on approximate variational
method with Gaussian proton profile (see Table I in \cite{szmaglinski06}) one 
observe systematically lower values resulting from the present method.
 For example:  AV14+UVII - 0.789 (old) and  0.745 (new),  UV14+TNI - 0.731 and
(old) 0.608 (new)  in $\fmc$. As a conclusion one may say that the correction to
the threshold density  is of the order of 10\%.  It seems natural that new
values of $n_{loc}$ are a little smaller since in our calculations both proton
wave function and neutron background present exact solutions of assumed
equations.

\begin{table}
\caption{Various parameters above  the threshold on the proton localization.}
\begin{tabular}{|l|c|c|c|c|c|}
\hline
& $n_n^\infty \rm~[fm^{-3}]$ & $n ~[\fmc]$ & $\Delta E/A \rm~ [MeV]$ 
& $E_p$ \rm~[MeV]& $\langle r_p \rangle [\fm]$ 
  \\ \hline \hline
 \parbox[t]{4mm}{\multirow{5}{*}{\rotatebox[origin=c]{90}{SLy4}}}
&0.785 & 0.805 & 0 & -45.99 & 0.674\\
&0.879 & 0.905 & -1.55 & -84.63 &     0.606\\
&0.972 & 1.007 & -3.73 & -131.18 & 0.548\\
&1.066 &     1.110 & -6.75 & -185.01 & 0.500\\
&1.160 & 1.214 & -10.81 & -245.55 & 0.459\\ \hline
 \parbox[t]{4mm}{\multirow{5}{*}{\rotatebox[origin=c]{90}{AV14+UVII}}}
&0.745 & 0.763 & 0 & -33.97 & 0.747\\
&0.861 & 0.885 & -3.52 & -114.48 &     0.624\\
&0.978 & 1.005 & -8.41 & -230.13 & 0.528\\
&1.094 &     1.123 & -13.44 & -376.14 & 0.456\\
&1.210 & 1.239 & -16.98 & -550.70 & 0.404\\ \hline
 \parbox[t]{4mm}{\multirow{5}{*}{\rotatebox[origin=c]{90}{UV14+TNI}}}
&0.610& 0.608& 0 & -23.62& 1.023\\
&0.725& 0.723& -0.22& -71.83&     0.826\\
&0.840& 0.838& -0.50& -137.2& 0.660\\
&0.955& 0.954& -0.80& -213.0& 0.564\\
&1.070& 1.069& -1.08& -297.0 & 0.500 \\ \hline
\hline
\end{tabular} 
\label{tab1}
\end{table}

\begin{figure}[t]
\includegraphics[clip,width=\columnwidth]{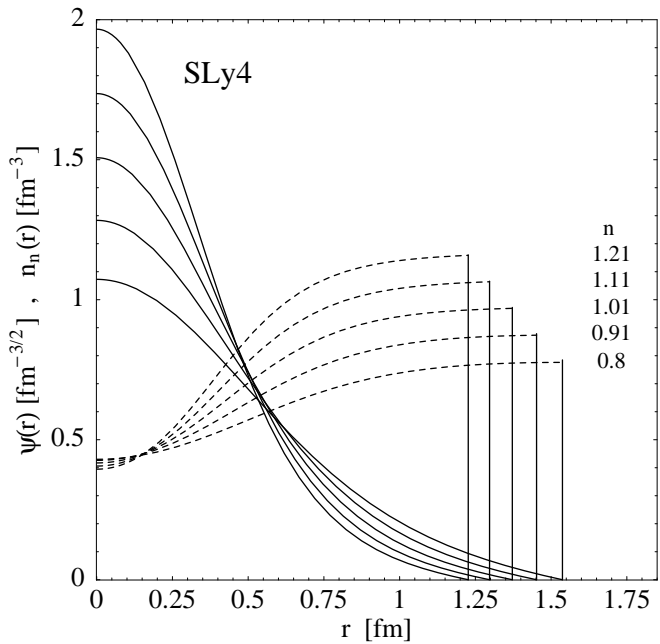}
\caption{The evolution of proton wave function (solid) and neutron background
(dashed) with
increasing mean baryon density $n$ for the SLy4 nuclear model. 
Vertical lines indicate the position of $R_{WS}$.}
\label{psi-evol-apr.eps}
\end{figure}

\begin{figure}[h]
\includegraphics[clip,width=\columnwidth]{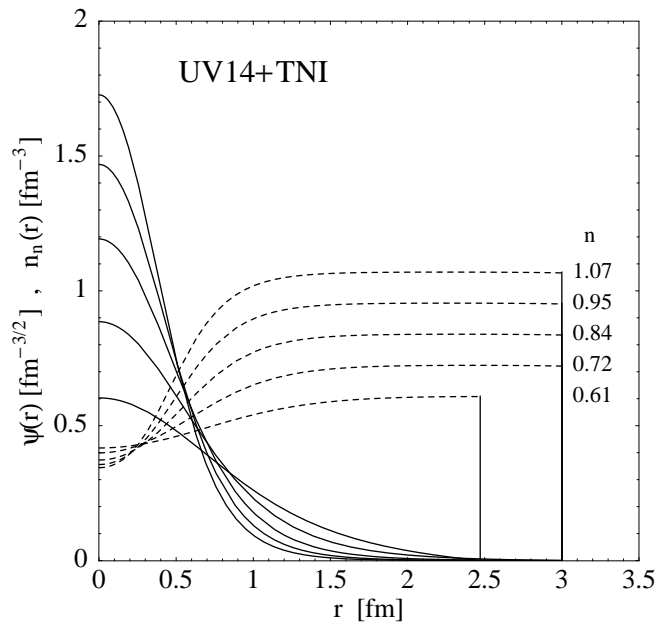}
\caption{The same like in the Fig.\ref{psi-evol-apr.eps} for the
 UV14+TNI nuclear model. For densities above $0.7~\fmc$ the $R_{WS}$ was greater
than $3~\fmc$.}
\label{psi-evol-uv14.eps}
\end{figure}

 The two models  (SLy4, AV14+UVII)  were analyzed in the whole range of
available  density: from the  threshold for localization (see first row for a
particular model in the Table \ref{tab1}) to the maximum density which is determined
by the maximum neutron star mass (the squares and full dots
in the  Fig.~\ref{x.eps}). In the case of the  third one (UV14+TNI),
 the range  of density relevant for localization was between the threshold and
the point where the protons disappear that means $x = 0$. It occurs
for $n=1.07~\fmc$. In the Fig.\ref{psi-evol-apr.eps} and
Fig.\ref{psi-evol-uv14.eps} the evolution with the baryon density of  proton
wave function and neutron background distribution is shown.  The vertical lines
indicate the W-S cell
radius $R_{WS}$. For the SLy4 model the $R_{WS}$ takes the smallest values 
 which
means that the cell contains about ten  neutrons.
 For the rest of the models, the $R_{WS}$ is greater and W-S cell
 containers from  10 to
several hundred of neutrons which justifies the
description of neutrons  by its local density $n_n(r)$. 

The behaviour of the proton wave function and neutron density is similar as in
the previous approximate calculations
\cite{kutschera90,kutschera93,kutschera02}. The proton mean radius $\langle r_p
\rangle$ decreases whereas the depth of the neutron well increases with the mean
density  of matter. The particular values of quantities  relevant for the proton
localization are shown in the Table \ref{tab1}. First two columns present the
neutron density $n_n^\infty$ at the W-S boundary Eq.~(\ref{nninf}) and the
mean baryon density $n$. The $\Delta E/A$ is the energy difference between the
homogeneous matter and  the state with localized proton Eq.~(\ref{DE}) taken per
total number of baryon in the cell. The $E_p$ presents  the proton energy
eigenvalue. For all models  the localization energy $\Delta E/A$ increases with
the density and the same happens to the proton energy $E_p$, so one may conclude
the proton is localized stronger  at higher densities. An interesting  fact is
that, in  case of  UV14+TNI model, although the proton fraction is very small, 
the strength  of localization, measured by the energy difference $\Delta E/A$ 
takes the  smallest values in comparison to the other models.

\vskip2.0cm

\section{Summary} In the present work we have solved the Lagrange-Euler
equations for a proton impurity with the extended Thomas-Fermi approach for
neutron background. The proton was treated as quantum particle immersed in the 
quasi-classical neutron sea. In a previous work the proton abundance was
assumed  to be infinitely small, e.i. the Wigner-Seitz cell was infinitely
large, $R_{WS}\rightarrow \infty$. Here we kept finite $R_{WS}$ determined by
the proton fraction which is fixed by the $\beta$-equilibrium occurring in neutron
star matter.  By minimization of the energy in the finite-size Wigner-Seitz
cell  we found exact solution for proton wave function and neutron background.
It turn out that proton localization still occurs for all the presented models. 
The localization threshold is slightly lower than in the previous work  where
the one-parameter method for energy minimization was used  \cite{kutschera02}.
In this work we have  investigated, in a rigorous way, the earlier ideas of the 
proton localization and have shown the phenomenon is plausible and worth further
research. Interesting issues involve: the crystallization of protons as has been
shown earlier \cite{kutschera95}, the influence of temperature  on the neutron
threshold density for proton localization \cite{skw14} and consequences for
neutron star cooling \cite{baiko99,baiko00}.

\section*{ACKNOWLEDGMENTS}

We are grateful to Marek Kutschera and Adam Szmagli\'{n}ski for helpful feedback
at the early stage of this work and fruitful discussions.

\end{document}